\documentclass[3p,times]{elsarticle}
\usepackage{amssymb}
\usepackage{graphicx}
\usepackage{amsmath}
\usepackage{color}
\usepackage[dvipsnames]{xcolor}
\usepackage{psfrag}
 \usepackage{lineno}
 \usepackage{amsthm}
 \usepackage{bm}
 \usepackage{mathtools}
  \usepackage{pdfpages}
\usepackage{url}

\usepackage{tikz}

\begin{document}

\begin{frontmatter}

\title{The quadratic map and its temporal and spectral properties}

\author{Rafael Alves da Costa$^1$ and Marcio Eisencraft}
\address{Telecommunication and Control Engineering Department, Escola Polit\'{e}cnica, University of S\~{a}o Paulo}
\cortext[cor]{Corresponding author: $^1$ rcosta@usp.br}

\begin{abstract}
This work numerically examines the temporal and spectral properties of a quadratic map. The quadratic map described in this study has quadratic non-linearity, and its theoretical analysis poses a challenge. Additionally, this map can be used for a fixed parameter value in chaos-based communication systems. Therefore, it is important to understand and, if possible, control the Power Spectral Density (PSD) generated by its signals. In practical communication systems, the bandwidth is limited, making it crucial to understand the spectral formatting of the employed signals.
\end{abstract}

\end{frontmatter}

\section{Introduction}

In this paper numerical results of the temporal and spectral characteristics of orbits generated by the quadratic map are presented. The quadratic map has quadratic nonlinearity and its theoretical analysis is challenging. Here, a numerical analysis was performed in order to obtain the Autocorrelation Sequence (ACS) and the Power Spectral Density (PSD) of this map.
In Section \ref{Sec::map}, the quadratic map is defined and its main properties are described. In Section \ref{acs_psd}, the ACS and PSD of this map are obtained numerically, and in Section \ref{Sec::concl}, we concisely deliver our conclusions.

\section{The quadratic map}\label{Sec::map}

The quadratic map, a modification of the logistic map \cite{May1976}, is defined by
\begin{equation}\label{eq::mapa}
	s(n+1) = f_Q\left(s(n)\right),
\end{equation}
in which
\begin{equation}\label{eq_quadratico}
f_Q(s)=-\frac{a}{2}s^2+\frac{a-2}{2},
\end{equation}
where $a\in \left[0,4\right]$ is a constant parameter and the initial condition $s(0)\in U=\left[-1,1\right]$. 

\begin{figure}[htb]
	\centering
\includegraphics[width=.7\columnwidth]{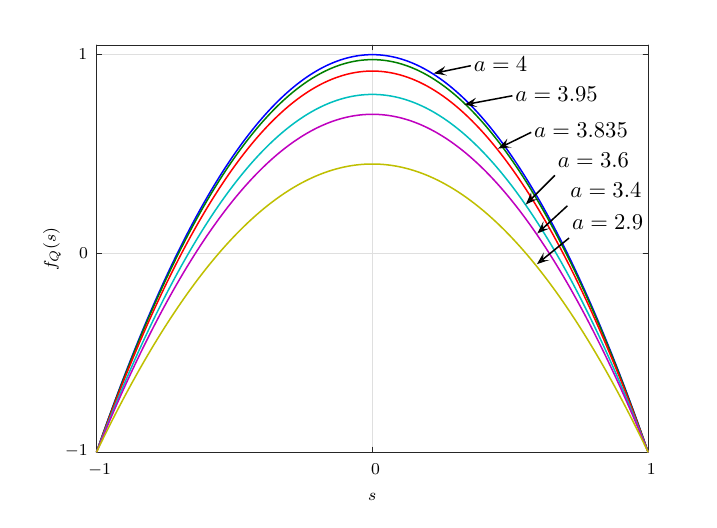}
	\caption{Quadratic map $f_Q(s)$ for different values of $a$.}	\label{example_maps}
\end{figure}

Fig. \ref{example_maps}, we present graphs of $f_Q(s)$ for a variety of values of $a$. These represent parabolas with their range in the interval from $\left [-1,\frac{a-2}{2}  \right ]$. Only for $a=4$ does the image equate to the domain $[-1,1]$.

\begin{figure}[ht]
      	\centering
      	\includegraphics[width=.72\columnwidth]{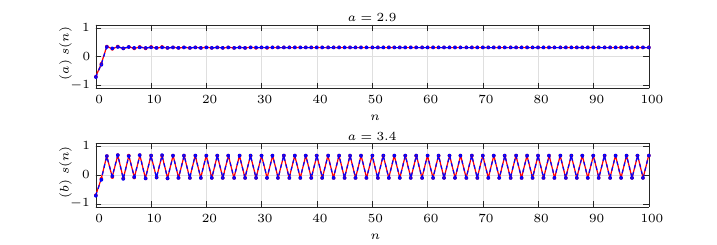}
      	\includegraphics[width=.72\columnwidth]{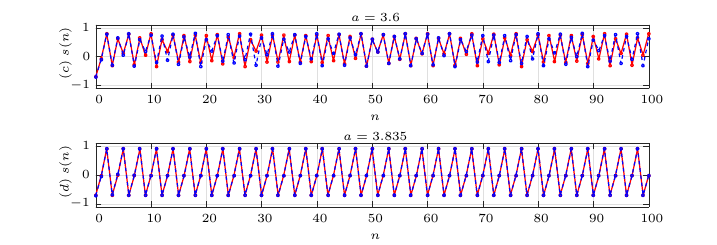}
      	\includegraphics[width=.72\columnwidth]{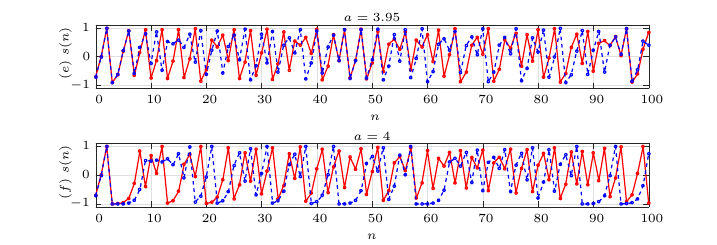}
      	\caption{Orbits of the quadratic map $f_Q(s)$ with $s(0)=-0.7$ (solid line) and $s(0)=-0.71$ (dashed line) for values of (a) $a=2.9$, (b) $a=3.4$, (c) $a=3.6$, (d) $a=3.835$, (e) $a=3.95$ and (f) $a=4$.}   
      	\label{LogisticoMapaOrb2}
      \end{figure}

In Figure \ref{LogisticoMapaOrb2}, orbits of the quadratic map $f_Q(s)$ with initial conditions $s(0)=-0.7$ and $s(0)=-0.71$ are shown for different values of $a$. As observed in Figure \ref{LogisticoMapaOrb2}(a) with $a=2.9$, with a few iterations, the two orbits converge to a fixed point. In Figure \ref{LogisticoMapaOrb2}(b) with $a=3.4$, after several iterations, the two orbits converge to a periodic orbit of period $2$. In Figure \ref{LogisticoMapaOrb2}(c) for $a=3.6$, there is evidence of chaotic behavior, that is, aperiodicity and sensitive dependence on initial conditions (SDIC). For $a=3.835$, in Figure \ref{LogisticoMapaOrb2}(d), the appearance of a periodic orbit, in this case of period 3, is noted. Finally, for $a=3.95$ and $a=4$, further evidence of chaotic behavior is observed in Figures \ref{LogisticoMapaOrb2}(e) and (f), respectively.

\begin{figure}[!htb]
  	\centering
  	\includegraphics[width=.86\columnwidth]{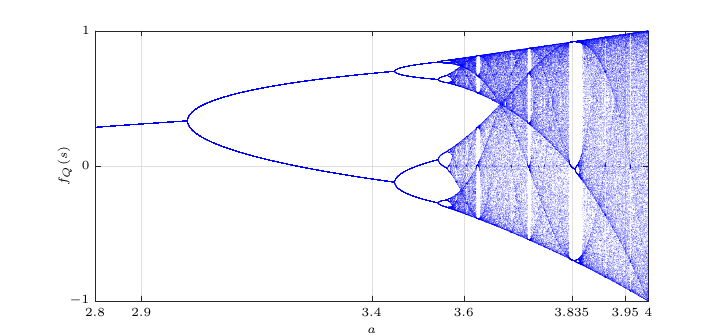}
  	\caption{Bifurcation diagram of the quadratic map $f_Q(s)$.} 
  	\label{LogisticoMapaOrb}
  \end{figure}

In Figure \ref{LogisticoMapaOrb}(a), the bifurcation diagram of \eqref{eq::mapa} is presented. In these simulations, $10^4$ values of $a$, equally spaced in the interval $(2.8,4)$, were used. For each value of $a$, an orbit was generated with $10^4$ points and an initial condition randomly chosen from the domain $U$. Only the final 70 points of the orbits are displayed in an attempt to eliminate transient behavior and capture the asymptotic solution. The values of $a$ used in Figure \ref{LogisticoMapaOrb} are indicated. The orbits converge to a fixed point for $0<a<3$. For $a>3$, a cascade of bifurcations appears with orbits of periods $2$, $4$, $8$,$\ldots$ up to $a=3.56995\ldots$. From this value of $a$ onwards, aperiodic behavior typical of chaotic signals emerges, with windows of periodicity. The largest of these, of period 3, appears in the interval $3.8284\ldots\leq a \leq3.8415\ldots$ \cite{Strogatz2001}.

As shown in the bifurcation diagram in Figure \ref{LogisticoMapaOrb}, the orbits of the map $f_Q(s)$ for different values of $a$ exhibit distinct asymptotic behaviors. Thus, the points that make up each trajectory for each $a$ are distributed in the domain $U$ in different ways.

\begin{figure}[htb]
	\begin{minipage}[b]{0.5\linewidth} 
		\centering
		\includegraphics[width=\columnwidth]{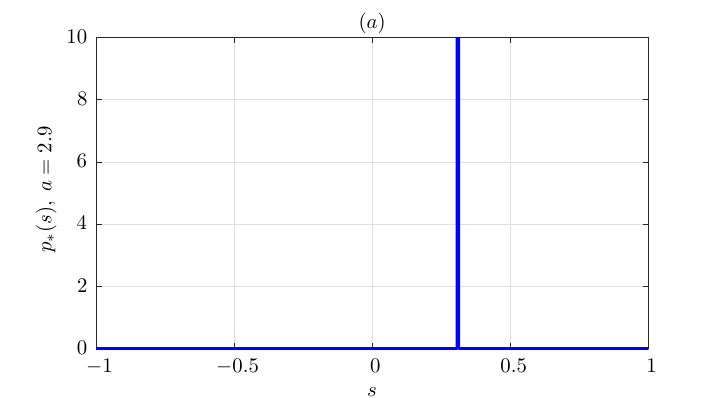}
		\includegraphics[width=\columnwidth]{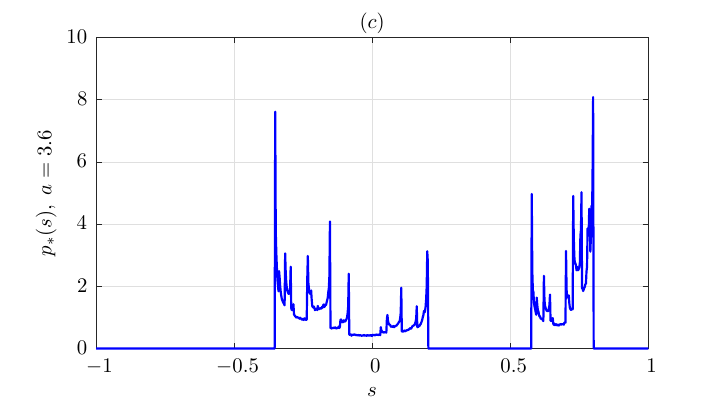}
		\includegraphics[width=\columnwidth]{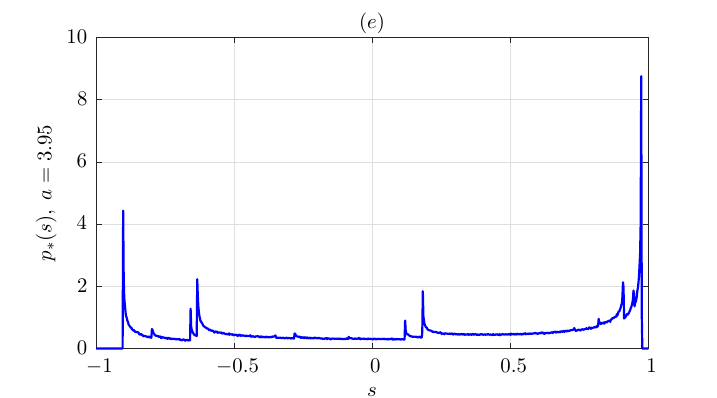}
	\end{minipage}
	\hspace{0.12cm} 
	\begin{minipage}[b]{0.5\linewidth}
		\centering
		\includegraphics[width=\columnwidth]{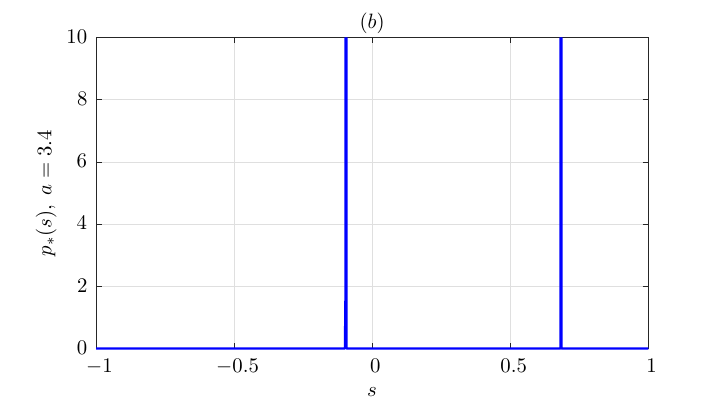}
		\includegraphics[width=\columnwidth]{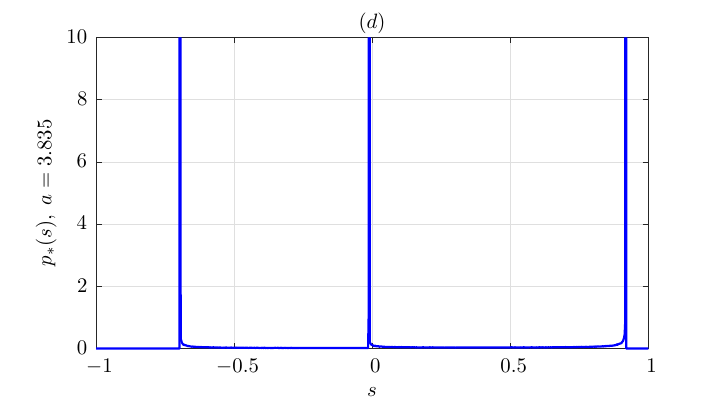}
		\includegraphics[width=\columnwidth]{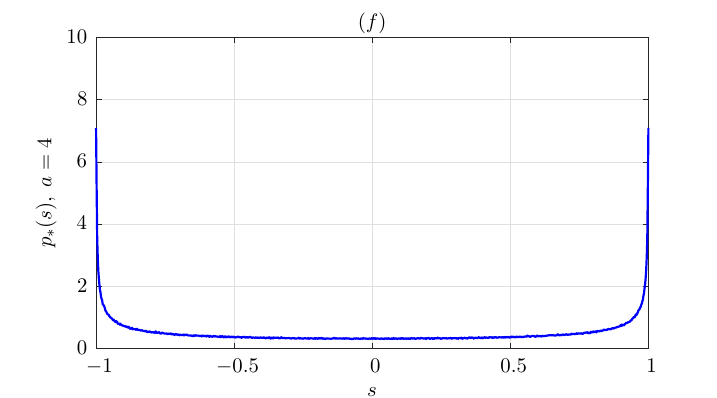}
	\end{minipage}
	\caption{Invariant density for the quadratic map $f_{Q}(s)$ for various values of $a$.}
	\label{LogisticoDensidade}
\end{figure}

In Figure \ref{LogisticoDensidade}, the invariant densities obtained through histograms for orbits of the quadratic map for different values of $a$ are shown. The histograms were generated from the iteration of the map $f_Q(\cdot)$ applied to a set of initial conditions uniformly distributed in the domain. For each value of $a$, $2\times 10^6$ initial conditions were taken, uniformly distributed in the domain $U= (-1,1)$, and the map was successively applied to this set of points until its distribution remained constant, that is, until its density became invariant \cite{Lasota1985}.

In Figures \ref{LogisticoDensidade}(a), (b), and (d), the invariant densities for $a = 2.9$, $a=3.4$, and $a=3.835$ are presented, respectively. As expected from Figure \ref{LogisticoMapaOrb}, the frequency of the map's points concentrates on one point for $a=2.9$, two points for $a=3.4$, and three points for $a=3.835$. Meanwhile, in Figures \ref{LogisticoDensidade}(c), (e), and (f) for $a=3.6$, $a=3.95$, and $a=4$, respectively, the distribution of the frequency of the points of $f_Q(\cdot)$ is not uniform, splitting into two bands within the domain $U$ for $a=3.835$, into a band that does not fill the entire map domain for $a=3.95$, and for $a=4$, the frequency of the points extends across the entire range of the domain $U$.

It can be shown that for $a=4$ the map $f_Q(s)$ has invariant density \cite{hao1989elementary},
\begin{equation}
\label{eqn_5_5}
p_\ast(s) = 
\begin{cases} 
\frac{1}{\pi\sqrt{1-s^2}}, & -1\leq s\leq 1\\
0, & \text{otherwise}
\end{cases}.
\end{equation}

\begin{figure}[!htb]
  	\centering
  	\includegraphics[width=.95\columnwidth]{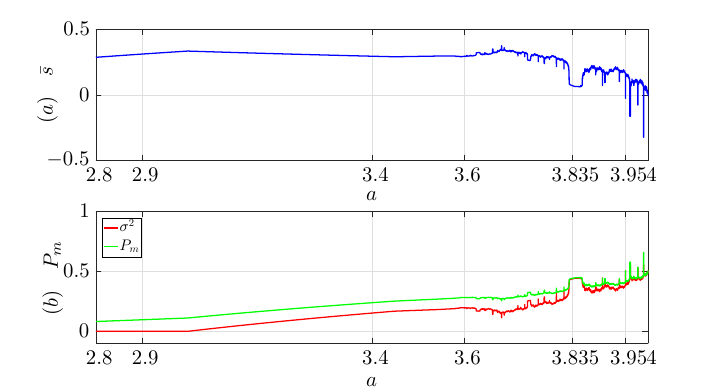}
  	\caption{ (a) Average value, (b) variance and average power of the signals generated by the quadratic map.}
  	\label{valormedio}
  \end{figure}

In Figure \ref{valormedio}(a), the average value of an orbit of the quadratic map is shown as a function of the parameter $a$.
In this simulation, $10^4$ values of $a$ were used, equally spaced in the interval $(2.8,4)$, and for each $a$ an orbit with $10^4$ points was generated. To capture the asymptotic behavior, the first $10^3$ points were eliminated.

Note that the signal generated by this map generally has a non-zero mean. In Figure \ref{valormedio}(b), the variance $\sigma^2$ and the average power $P_m=\sigma^2+\bar{s}^2$ for an orbit of the quadratic map are shown.

In particular, for $a=4$, it is possible to analytically calculate the mean, variance, and average power of the orbits. In this case, using \eqref{eqn_5_5}.
\begin{align}
\label{4} 
\bar{s} &= \int_{-1}^{1}s\frac{1}{\pi\sqrt{1-s^2}}ds =0
\end{align}
and the average power
\begin{align}
\label{5} 
P_m&=\int_{-1}^{1}s^2\frac{1}{\pi\sqrt{1-s^2}}= \frac{1}{2}.
\end{align}

\section{Temporal and Spectral Characterization}\label{acs_psd}

\begin{figure}[!htb]
	\centering
	\includegraphics[width=\columnwidth]{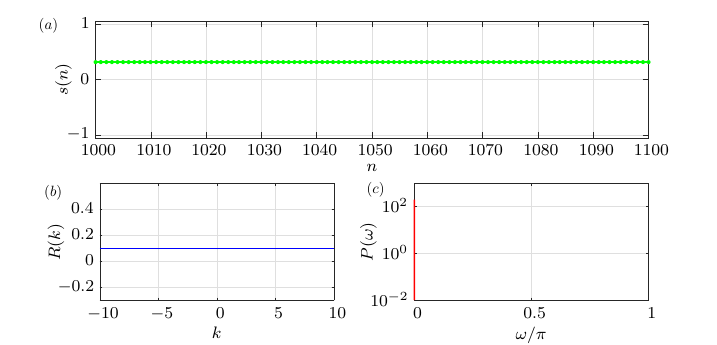}
	\caption{(a) Orbit, (b) ACS, and (c) PSD of the map $f_{Q}(\cdot)$ for $a=2.9$.}
	\label{SACDEP1}
\end{figure}

\begin{figure}[!htb]
	\centering
	\includegraphics[width=\columnwidth]{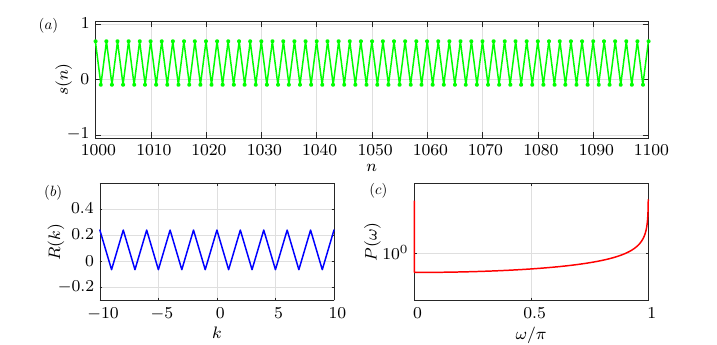}
	\caption{(a) Orbit, (b) ACS, and (c) PSD of the map $f_{Q}(\cdot)$ for $a=3.4$.}
	\label{SACDEP2}
\end{figure}

\begin{figure}[!htb]
	\centering
	\includegraphics[width=\columnwidth]{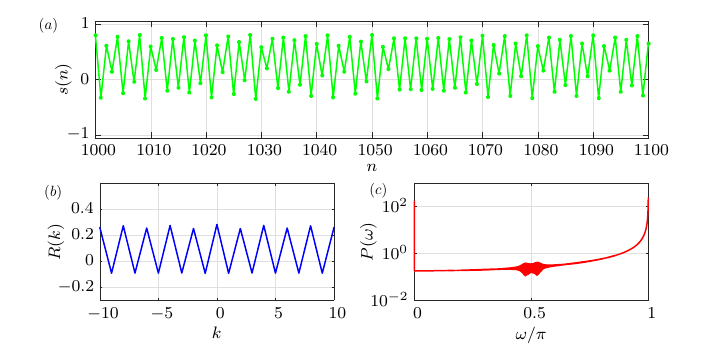}
	\caption{(a) Orbit, (b) ACS, and (c) PSD of the map $f_{Q}(\cdot)$ for $a=3.6$.}
	\label{SACDEP3}
\end{figure}

\begin{figure}[!htb]
	\centering
	\includegraphics[width=\columnwidth]{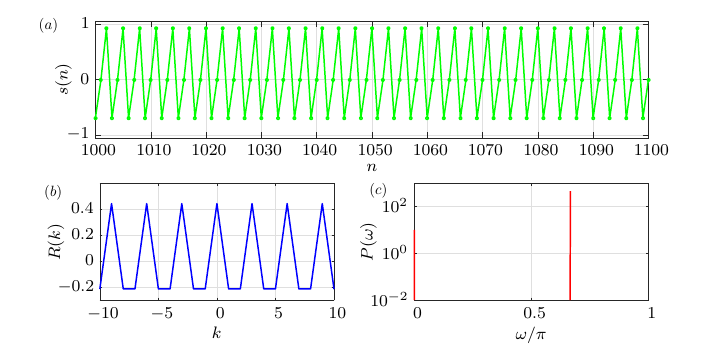}
	\caption{(a) Orbit, (b) ACS, and (c) PSD of the map $f_{Q}(\cdot)$ for $a=3.835$.}
	\label{SACDEP4}
\end{figure}

\begin{figure}[!htb]
	\centering
	\includegraphics[width=\columnwidth]{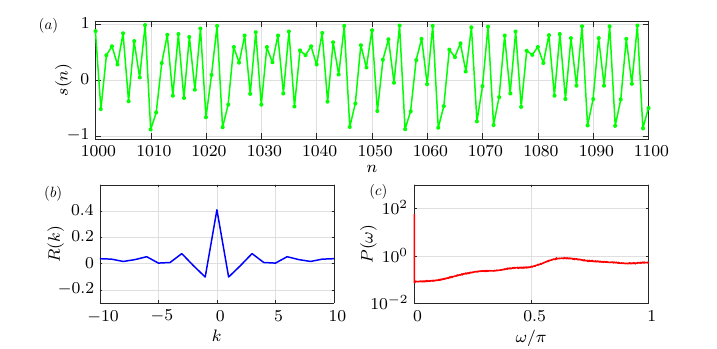}
	\caption{(a) Orbit, (b) ACS, and (c) PSD of the map $f_{Q}(\cdot)$ for  $a=3.95$.}
	\label{SACDEP5}
\end{figure}

\begin{figure}[!htb]
	\centering
	\includegraphics[width=\columnwidth]{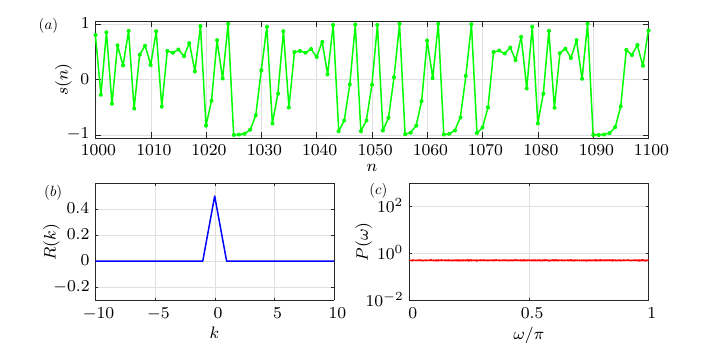}
	\caption{(a) Orbit, (b) ACS, and (c) PSD of the map $f_{Q}(\cdot)$ for  $a=4$.}
	\label{SACDEP6}
\end{figure}

In Figures \ref{SACDEP1} to \ref{SACDEP6}, plots of an orbit and estimates of the ACS and PSD for various values of $a$ are shown.

Orbits were generated for each $a$ with $5\times 10^3$ points and the first $10^3$ points were eliminated in an attempt to capture the asymptotic stretch. In the ACS and PSD simulations, the average of $500$ sample functions with initial conditions uniformly distributed in the domain $U$ was taken.

Figure \ref{SACDEP1}(a) shows a section of the orbit of the map $f_{Q}(s)$, for $a=2.9$. As seen in the previous section, a constant signal $s(n)\approx 0.31$ is obtained for sufficiently large $n$. In addition, the ACS is also constant and the PSD has a single DC component, respectively, in Figures \ref{SACDEP1}(b) and (c).

In Figure \ref{SACDEP2}(a), a section of the quadratic map orbit is illustrated for $a=3.4$. As seen in Figure \ref{LogisticoMapaOrb}(a), in this case, the asymptotic solution for the map is a period-2 orbit. In Figure \ref{SACDEP2}(b) and (c), the ACS oscillates between two values, and the PSD has a DC component and a spectral component at $\omega = \pi$, which is the signal's frequency.

In Figure \ref{SACDEP3}(a), the case $a=3.6$ is illustrated. As seen earlier, in this case, the generated orbit is chaotic but contains pronounced periodic components of period 2 and 4, which are reflected in the larger delays of the ACS at $k=2$ and $k=4$, and in the PSD by the frequencies $\omega = \pi$ and $\omega = \frac{\pi}{2}$ shown in Figures \ref{SACDEP3}(b) and (c), respectively.

In Figure \ref{SACDEP4}(a), for $a=3.835$, a section of the period-3 orbit is shown. It can be seen in Figure \ref{SACDEP4}(b) and (c) that the ACS is periodic with period 3 and the PSD, as expected, resulted in an impulse at $\frac{2\pi}{3}$.

In Figure \ref{SACDEP5}(a), a section of the chaotic signal generated for $a=3.95$ is shown. In Figure \ref{SACDEP5}(b), the ACS has oscillations that decay with $k$, and in Figure \ref{SACDEP5}(c), the PSD exhibits the behavior of high-pass signals, and a DC component can be observed.

In Figure \ref{SACDEP6}(a), a section of the orbit of the map $f_Q(s)$ for $a=4$ is displayed. In Figure \ref{SACDEP6}(b), an impulsive behavior of the ACS is observed. Lastly, in Figure \ref{SACDEP6}(c), it is noted that the spectrum is flat, that is, the power is equally distributed across all frequencies.

The ACS can be obtained analytically in the case of $a=4$. For this value of $a$, it is possible to derive a closed-form for the $n$-th sample of $s(n)$ \cite{Alligood2000}
\begin{align}
\label{8}
s(n) &= -\cos\left ( 2^n\arccos \left ( -s(0) \right ) \right ).
\end{align}

Unlike the maps $f_I(\cdot)$ \cite{Eisencraft2010} and $f_{B}(\cdot)$ \cite{Costa2017a}, which have uniform invariant density $p_*(\cdot)$, the quadratic map for $a=4$ has non-uniform invariant density given by \eqref{eqn_5_5}. Thus, for this value of $a$, using  \eqref{eqn_5_5} and \eqref{8}, it can be shown \cite{Costa2017a} that $R(k)$ can be written as
\begin{align}
\label{9}
R(k) &= \int_{-1}^{1}\frac{-x\left ( \cos \left ( 2^k\arccos \left ( -x \right ) \right ) \right )}{\pi\sqrt{1-x^2}}dx.
\end{align}

Rewriting $\cos \left ( 2^k\arccos \left ( -x \right ) \right )$ as $\cos \left ( 2^k\left ( \pi-\arccos \left ( x \right ) \right ) \right )$ and using the identities $\cos \left ( \alpha-\beta \right )=\cos\left ( \alpha \right )\cos\left ( \beta \right )+\sin \left ( \alpha \right )\sin\left ( \beta \right )$, with $\sin\left ( 2^k\pi \right )= 0$, and
Rewriting equation \eqref{19} as

\begin{equation}
\label{19}
\cos(2^k\pi) = \begin{cases}
-1,& k=0\\
1, & k>0
\end{cases},
\end{equation}
we can infer that for $k=0$, we have
\begin{align}
\label{7}\nonumber
R(0)&=\int_{-\infty }^{\infty }x\cos \left ( \arccos \left ( x \right ) \right )p_{\ast}(x)dx \\ 
&= \int_{-1}^{1}\frac{x^2}{\pi\sqrt{1-x^2}}dx=\frac{1}{2}
\end{align}
and for $k> 0$,
\begin{align}
\label{16}
R(k) &=\frac{1}{\pi } \int_{-1}^{1}\frac{-x\left ( \cos \left ( 2^k\arccos \left ( x \right ) \right ) \right )}{\sqrt{1-x^2}}dx.
\end{align}

Note that in \eqref{16}, the numerator features a product of an odd function and an even function, resulting in an odd function. The denominator, meanwhile, is an even function. Hence, the integrand as a whole is an odd function. Since the integral is calculated over a symmetric interval relative to the origin, we can conclude that its result will be zero. From the symmetry of $R(k)$ for a real signal, we then get,
 \begin{align}
 \label{18}
R(k) &=  \frac{1}{2}\delta(k)
 \end{align}
 and
 \begin{equation}
\label{2_41}
P(\omega)=\sum_{k=-\infty }^{\infty }R(k)e^{-j\omega k}=\frac{1}{2}.
\end{equation}

Therefore, in spectral terms, the signals generated by the quadratic map with $a=4$ are equivalent to sample functions of white noise.

\begin{figure}[!htb]
	\centering
	\includegraphics[width=.85\columnwidth]{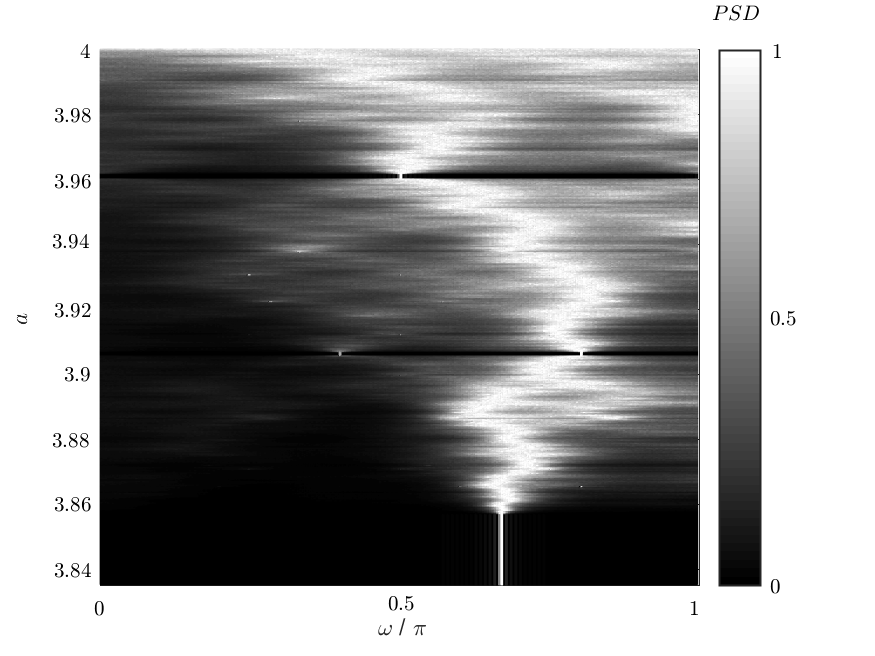}
	\caption{PSD of the map $f_{Q}(\cdot)$ for $3.835\leq a\leq4$.}
	\label{DEPQuadra}
\end{figure}

Analytical calculations for $a\neq4$ are prohibitive as there are no closed-form solutions for the invariant densities, which have very complicated shapes, as seen in Figure \ref{LogisticoDensidade}. Therefore, in Figure \ref{DEPQuadra} we present the PSD of the map $f_Q(\cdot)$ in the interval  $3.835\leq a\leq4$, excluding the DC component. It is observed that for the parameter $a$ in the analyzed interval, the PSD is predominantly narrowband and concentrated at high frequencies, being equivalent to white noise only when $a = 4$.
 
\section{Conclusions}\label{Sec::concl}

In this work, the temporal and spectral characteristics of the quadratic map were studied numerically.
For $a\neq4$, obtaining the ACS and PSD analytically is more challenging compared to the analyses of previous works \cite{Eisencraft2010,Costa2017a,Costa2019}, due to the invariant density not being uniform and not having a closed form expression.

\section*{Acknowledgments}

This work was partially supported by the
National Council for Scientific and Technological Development (CNPq-Brazil) (grant number 311039/2019-7) and by the Coordination for the Improvement of Higher Education Personnel (CAPES-Brazil) (grant number 001).

\bibliographystyle{elsarticle-num}

\bibliography{Biblio-Marcio}

\begin{thebibliography}{10}
\expandafter\ifx\csname url\endcsname\relax
  \def\url#1{\texttt{#1}}\fi
\expandafter\ifx\csname urlprefix\endcsname\relax\def\urlprefix{URL }\fi
\expandafter\ifx\csname href\endcsname\relax
  \def\href#1#2{#2} \def\path#1{#1}\fi

\bibitem{May1976}
R. M. May, Simple mathematical models with very complicated dynamics, Nature 261 (5560) (1976) 459–467.

\bibitem{Strogatz2001}
S. H. Strogatz, Nonlinear Dynamics and Chaos, The Perseus Books Group, 2001.

\bibitem{Lasota1985}
A.~Lasota, M.~C. Mackey, Probabilistic Properties of Deterministic Systems,
  Cambridge University Press, 1985.

\bibitem{hao1989elementary}
B.-L. Hao, Elementary symbolic dynamics and chaos in dissipative systems, World Scientific, 1989.

\bibitem{Alligood2000}
K.~T. Alligood, T.~D. Sauer, J.~A. Yorke, Chaos, Textbooks in Mathematical
  Sciences, Springer New York, 2000.

\bibitem{Eisencraft2010}
M.~Eisencraft, D.~Kato, L.~Monteiro, Spectral properties of chaotic signals
  generated by the skew tent map, Signal Processing 90~(1) (2010) 385--390.
\newblock \href {https://doi.org/10.1016/j.sigpro.2009.06.018}
  {\path{doi:10.1016/j.sigpro.2009.06.018}}.


\bibitem{Costa2017a}
R.~A. Costa, M.~B. Loiola, M.~Eisencraft, Correlation and spectral properties
  of chaotic signals generated by a piecewise-linear map with multiple
  segments, Signal Processing 133 (2017) 187--191.
\newblock \href {https://doi.org/10.1016/j.sigpro.2016.10.025}
  {\path{doi:10.1016/j.sigpro.2016.10.025}}.

\bibitem{Costa2019}
R.~A. da~Costa, M.~Eisencraft, Spectral characteristics of a general piecewise
  linear chaotic signal generator, Communications in Nonlinear Science and
  Numerical Simulation 72 (2019) 441--448.
\newblock \href {https://doi.org/10.1016/j.cnsns.2019.01.002}
  {\path{doi:10.1016/j.cnsns.2019.01.002}}.


\end{thebibliography}

\end{document}